\documentclass[12pt]{article}
\usepackage{amsmath}
\usepackage{graphicx}
\usepackage{natbib}
\usepackage{url} % not crucial - just used below for the URL 
\usepackage{hyperref}

\newcommand{\blind}{0}

\begin{document}

\def\spacingset#1{\renewcommand{\baselinestretch}%
{#1}\small\normalsize} \spacingset{1}

%%%%%%%%%%%%%%%%%%%%%%%%%%%%%%%%%%%%%%%%%%%%%%%%%%%%%%%%%%%%%%%%%%%%%%%%%%%%%%

\if0\blind
{
  \title{\bf Network Layout Algorithm with Covariate Smoothing}
  \author{Octavious Smiley* 
 \\
    Till Hoffmann* \\
    and \\
    Jukka-Pekka Onnela*\\
    Department of Biostatistics, Harvard T.H. Chan School of \\ Public Health, Boston, MA*}
  \maketitle
} \fi

\if1\blind
{
  \bigskip
  \bigskip
  \bigskip
  \begin{center}
    {\LARGE\bf Title}
\end{center}
  \medskip
} \fi

\bigskip
\begin{abstract}
Network science explores intricate connections among objects, employed in diverse domains like social interactions, fraud detection, and disease spread. Visualization of networks facilitates conceptualizing research questions and forming scientific hypotheses. Networks, as mathematical high-dimensional objects, require dimensionality reduction for (planar) visualization. Visualizing empirical networks present additional challenges. They often contain false positive (spurious) and false negative (missing) edges. Traditional visualization methods don't account for errors in observation, potentially biasing interpretations.
Moreover, contemporary network data includes rich nodal attributes. However, traditional methods neglect these attributes when computing node locations. Our visualization approach aims to leverage nodal attribute richness to compensate for network data limitations. We employ a statistical model estimating the probability of edge connections between nodes based on their covariates. We enhance the Fruchterman-Reingold algorithm to incorporate estimated dyad connection probabilities, allowing practitioners to balance reliance on observed versus estimated edges. We explore optimal smoothing levels, offering a natural way to include relevant nodal information in layouts. Results demonstrate the effectiveness of our method in achieving robust network visualization, providing insights for improved analysis.
\end{abstract}

\noindent%
\vfill

\newpage
\spacingset{1.75} % DON'T change the spacing!
\section{Introduction}

\section{Introduction} 

Graph visualizations have played an important role in generating new insights and hypotheses in network science \citep{Viegas, Didimo, FR, Mcgee, Jacomy}. Networks (graphs) are mathematically high dimensional objects, and any (planar) visualization method requires reducing their dimensionality to two. Given the massive reduction in dimensionality, it is not surprising that there are many ways to visualize graphs. 

The literature abounds with interesting and insightful network visualizations. One such example is demonstrated by \citet{Viegas}. They visualized social networks utilizing emails as a proxy for relationships in two separate ways. They viewed both the traditional network with the email contacts as nodes and a graph depicting the amount of emails exchanged over time between ego and each different contact, showing large concentrations of interaction during certain periods as well as times when almost no emails were exchanged. After observing and interviewing the users of the systems, the authors concluded that, when used in tandem, these visualizations complemented and clarified each other’s depiction of a person’s social network \citep{Viegas}. \citet{Didimo} noted the economic impact of financial crimes such as money laundering and fraud on global economies. Governmental attempts to thwart such crimes have included large scale tracking of financial data. \citet{Didimo}recognized that the large volume and complexity of collected data can be modeled as a financial activity network (FAN) and noted consensus that analysis of these data would benefit from network visualization tools. To help alleviate the burden of financial crimes, the authors have described a new software system for the analysis of FANs that is designed to support the analysts in the discovery of criminal patterns and utilizes visual interactive tools combined with ad-hoc clustering techniques and customizable layout managements \citep{Didimo}. \citet{brandes} were interested in whether the characteristics of organizations in local drug policy could explain the difference in the availability of HIV/AIDS preventative measures. Data was gathered on the participating organizations in each municipality as well as on the relations between them. Closeness and betweenness centrality were calculated for informal communication and then visualized to analyze how the structural characteristics of the network could be linked to the outcomes in local drug policy. The nodes were placed according to their centrality score where the most central node is placed in the center of the drawing and the others with decreasing centrality toward the edges of the structure. Tying this network to HIV treatment outcomes, they believe this type of layout contributes to the analysis of questions ``Who has the power?'' and ``What are the consequences of the power structure?'' \citep{brandes}.

Inherent in visualizing network data is a need to conceptualize the randomness or ambiguities in the data and/or drawing algorithm \citep{Wang, Schulz, Ament, Guo, MacEachren}. \citet{Wang} illustrated that the value of visualizing network data has inspired many automated drawing algorithms that seek to optimize several aesthetic criteria such as reducing edge lengths, increasing visual overlap between community structures, and node/edge bundles. However, it is unlikely that a single layout can satisfy all criteria and there tends to be a trade off that can hinder interpretations of the graphs. \citet{Wang} proposed methods to analyze several aesthetic metrics to identify network and algorithm combinations that may lead to random ambiguities in a layout's interpretation to better guide which layout to present \citep{Wang}. \citet{Schulz} noted the confusion uncertainty in the underlying data and drawing algorithm can bring to one's interpretation. Their aim was to visualize the distribution of possible realizations of a probabilistic graph that reflects certainty and uncertainty equally well. Their results provide insights into probability distributions for the entire network beyond individual nodes or edges, reveals the general limitations of force-directed layouts, and allows the user to recognize that some features in a particular graph layout are simply there by unintended chance.

Furthermore, in standard force drawing algorithms, there tends to be sensitivity to the observed network \citep{FR, Kamada}. \citet{FR} developed a network layout algorithm (FR) with their heuristic focus on edge lengths. They utilized attractive and repulsive forces where all pairwise nodes have a repulsive force and pairs with an edge have an additional attractive force \citep{FR}. This network drawing algorithm provides pleasing aesthetics such as relatively short edge lengths and few edge crossings. However, it depends completely on the observed network and its accuracy. 

Network data, like any other data, is susceptible to inaccuracies during data collection \citep{Rosenblatt, Kossinets, Folch}. Hence, missing data and/or inaccuracies in the observed network may bias the graph visualization and thereby affect its interpretation. As the field has progressed, increasingly rich network data has become available, and it is now common that nodes have attributes (covariates) associated with them. Traditional visualization methods however do not consider nodal covariates when computing node locations.

In this paper, we propose a visualization framework that attempts to leverage the ``richness'' of the nodal attributes to compensate for the ``poorness'' (e.g., missing edges) of the network data.  At the core of our model-based approach to network visualization is the idea that we fit a simple statistical model that estimates the probability that any dyad (node pair) is connected with an edge given the covariates of the two nodes. We modified one of the standard network visualization methods, the Fruchterman-Reingold (FR) algorithm, to incorporate estimated dyad connection probabilities in a manner that enables the practitioner to interpolate anywhere between the two extremes of relying on observed edges only versus relying on estimated edge probabilities only. 

Our paper is organized as follows. In Section \ref{sec:methods}, we introduce our method, which relies heavily on the seminal work by \citet{FR} on force-directed graph layouts. Our method has one user-specified tuning parameter, which controls the relative importance of the observed network and covariates, and we offer a heuristic for selecting reasonable values for it. In Section \ref{sec:results}, we show the results of our method for both simulated and empirical networks. We conclude in Section \ref{sec:conclusion}.

\section{Methods}
\label{sec:methods}

\subsection{Energy Function}
Consider a simple undirected graph $G$ with $n$ nodes and a corresponding $n\times n$ adjacency matrix $\mathbf{A}$, where $A_{i j} = 1$ if there is an edge between nodes $i$ and $j$ and 0 otherwise. We take $\mathbf{X}$ to be our $n\times p$ nodal feature (covariate) matrix where $p$ is the number of features. We assume that the true but unknown network data generation mechanism is given by the true model with a corresponding $n\times n$ matrix of the true pairwise linkage probabilities $\tilde{\mathbf{B}}$. In other words, $\tilde{\mathbf{B}}_{i j} = p_{i j}$ implies there is a connection between nodes $i$ and $j$ with probability of $p_{i j}$ under the model. As such, we consider $G$ (and $\mathbf{A}$) to correspond to a single realization of $\tilde{M}$. While the network is generated by an unknown model $\tilde{M}$, we propose approximating the true model $\tilde{M}$ with a simpler dyadic model $M$ with the goal of regularizing the graph layout. The linkage probabilities corresponding to model $M$ are given by the matrix $\mathbf{B}$ which also is unknown and requires estimation. In this paper we construct an estimate $\mathbf{\hat{B}}$ of $\mathbf{B}$ using information from $\mathbf{X}$ and $\mathbf{A}$ such that $\mathbf{\hat{B}} = \mathbf{\hat{B}(X, A)}$.

Similar to FR, we bound our layout on a fixed domain \citep{FR}. Let $\mathbf{P}$ be an $n\times 2$ matrix containing all of the planar embedding coordinates of the nodal positions on a bounded layout. Then $\mathbf{D}$ = $\mathbf{D(P)}$ is an $n\times n$ matrix containing the pairwise distances of the nodes of a particular layout $\mathbf{P}$ where $D_{i j}$ is the Euclidean distance between nodes $i$ and $j$. We represent the forces of the FR algorithm in terms of the energy function 
\begin{align}
Q_1 &= \sum_{i = 1}^{n} \sum_{j = 1}^{i - 1}\left( \frac{q^2}{D_{i j}} + kA_{i j}D_{i j}^2 \right),
\end{align}
where $q$ is the repulsion constant and $k$ is the attraction constant. We take $q$ = $k$ = $1$, such that 
\begin{align}
Q_1 &= \sum_{i = 1}^{n} \sum_{j = 1}^{i - 1} \left(\frac{1}{D_{i j}} + A_{i j}D_{i j}^2 \right).
\end{align}
However, other values can be investigated if there is interest to deeply study a variety of repulsion/attraction levels. The desired layout is chosen by minimizing $Q_1$ with respect to $\mathbf{D}$. To incorporate covariate smoothing in the FR algorithm, we use the modified energy function
\begin{align}
Q_2 &= \sum_{i = 1}^{n} \sum_{j = 1}^{i - 1} \left( \frac{1}{D_{i j}} + [(1 - \gamma)A_{i j} + \gamma \hat{B}_{i j}]D_{i j}^2 \right),
\end{align}
where $\gamma\in [0, 1]$ is fixed and is considered our smoothing parameter. This allows us to include the pairwise probabilities inside the attractive forces of our energy function. Again, the desired layout is chosen by minimizing $Q_2$ with respect to $\mathbf{D}$. $\gamma=0$ defaults to the standard FR algorithm and is interpreted here as ``no smoothing.'' However, $\gamma=1$ ignores the observed network completely, except through estimation of $\mathbf{\hat{B}}$, and is considered here as ``maximum smoothing.''

\subsection{Tuning Variable Selection}
\label{subsec:tuning}
We recognize there is flexibility in the choice of $\gamma$. Neither extreme of $\gamma=0$ or $\gamma=1$ is likely to be the best choice for the problem at hand. Here, we propose a metric that measures some of the aesthetic properties in simulated networks as a function of $\gamma$ and recommend $\gamma$ values for different network and nodal feature data. 

If $\mathbf{P_\text{$\gamma$}}$ is the matrix of nodal positions for fixed $\gamma$, then $\mathbf{P_\text{$\gamma$}} = \mathbf{P_\text{$\gamma$}(A, \hat{B})}$. Utilizing $\mathbf{P_\text{$\gamma$}}$ allows us to calculate the nodal distance matrix $\mathbf{D_\text{$\gamma$}}=\mathbf{D_\text{$\gamma$}(P_\text{$\gamma$})}$. We only consider simple undirected graphs, and we define $\bar{\hat{B}}$ and $\bar{D}_\text{$\gamma$}$ as the element-wise averages of the matrices B and D, respectively. We define $\mathbf{\hat{B}^C}$ and $\mathbf{D_\text{$\gamma$}^C}$ as centered versions of $\mathbf{\hat{B}}$ and $\mathbf{D_\text{$\gamma$}}$:
\begin{align}
%\mathbf{\hat{B}^C} &= \frac{\mathbf{\hat{B}} - \bar{\hat{B}}}{\hat{B}_{\sigma}} \\
%\mathbf{D_\text{$\gamma$}^C} &= \frac{\mathbf{D_\text{$\gamma$}} %\bar{D}_\text{$\gamma$}}{D_{_\text{$\gamma$, $\sigma$}}}. 
\mathbf{\hat{B}^C} &= \mathbf{\hat{B}} - \bar{\hat{B}} \\
\mathbf{D_\text{$\gamma$}^C} &= \mathbf{D_\text{$\gamma$}} - \bar{D}_\text{$\gamma$}.
\end{align}
We want to consider the smoothing effect as well as visual aesthetics in our selection. So, we further define
\begin{align}
m_{\gamma} &= \sum_{i j}  D_{ij \gamma}^C  \hat{B}_{i j}^C
\end{align}
to capture the smoothing effect and consider the standardization 
\begin{align}
\mathbf{m^S} &= \frac{\mathbf{m}- \bar{m}}{m_\sigma}.
\end{align}
Here the vector $\mathbf{m}$ is generated across a range of 20 ordered $\gamma$s sampled along the grid [0,1], $\bar{m}$ is the mean of $\mathbf{m}$, and $m_\sigma$ is the standard deviation of $\mathbf{m}$. For clarity, $\mathbf{m}$ = $[m_{\gamma_1}, m_{\gamma_2}, \ldots]$ implies $\gamma_1$ is the smallest of the sampled $\gamma$s, then $\gamma_2$, etc. We analyze the relative change in $\mathbf{m}$ over different $\gamma$ values, utilizing $\mathbf{m^S}$, where lower values are seen as better. This is because in our final matrix $\mathbf{P_\text{$\gamma$}}$, we expect nodal pairs with higher linkage probabilities to be near each other and pairs with lower linkage probabilities to be further apart. The transformations $\mathbf{\hat{B}^C}$ and $\mathbf{D^C}$ allow us to penalize nodal distance and probability pairs that are not consistent with this expectation. For instance, nodal distances that are much larger than the average distance that have corresponding probabilities much higher than the average probability will have a positive product and this is the same for the reversal where small distances correspond to small probabilities. However, small nodal distances that correspond to high linkage probabilities will have a negative contribution as well as large nodal distances that correspond to small linkage probabilities. Standardizing our metric also allows for us to combine it with other standardized metrics on an additive scale. By construction, $\gamma = 1$ corresponds to maximum smoothing, and in expectation, assuming our algorithm always finds the global minimum for our energy function that controls the layout, our metric $m_\gamma$ is monotone decreasing in $\gamma$. However, depending on the original network structure and linkage probabilities, aesthetically pleasing smoothing may reasonably be achieved at lower values of $\gamma$. Hence, we expect some scenarios where our metric reasonably plateaus after a particular threshold of $\gamma$ is reached. This is beneficial because we can expect smoothing in our layout without sacrificing the explicit information contained in the observed network structure, where some of the aesthetic benefits include shorter edge lengths. In addition to consider the aesthetics of the graph, we define
\begin{align}
e_{\gamma} &= \sum_{i j} D_{i j \gamma}^C A_{i j} 
\end{align}
to capture the changes in edge lengths as a function of $\gamma$ and again consider the standardization 
\begin{align}
\mathbf{e^S} &= \frac{\mathbf{e}- \bar{e}}{e_\sigma},
\end{align}
where the vector $\mathbf{e}$ is generated using the ordered sampled of $\gamma$s, $\bar{e}$ is the mean, and $e_\sigma$ is the standard deviation. Now we modify our original metric such that
\begin{align}
\psi_{\gamma_i} &= m_{\gamma_i}^S + e_{\gamma_i}^S,
\end{align}
where $\gamma_i$ corresponds to the $i^{th}$ ordered entry of the $\gamma$s from the grid and therefore the $i^{th}$ entry of $\mathbf{\psi}$, $\mathbf{m^S}$, $\mathbf{e^S}$. So, $\mathbf{\psi}$ is our metric used to select for $\gamma$. We interpret this metric as a measure of the relative change in smoothing among our $\gamma$s added together with a measure of the relative change in edge lengths among the $\gamma$s. As we increase $\gamma$, the smoothing metric $m_{\gamma}$ goes down, and the edge lengths $e_{\gamma}$ goes up in expectation, so we seek a value of $\gamma$ that provides the minimum value of $\mathbf{\psi}$. However, it is worth noting that the expected total edge length is monotone increasing in $\gamma$ due to the steady departure from the FR algorithm. Again, when we increase the smoothing parameter, we sacrifice more of the observed network structure when drawing the graph which results in $e_{\gamma}$ becoming larger as we move away from the FR. Utilizing $\mathbf{\psi}$ allows us to turn our $\gamma$ selection problem into an optimization problem as we seek to smooth our layout and optimize its impact on the graph aesthetics. To increase the accuracy of the metric, $\mathbf{\psi}$ can be generated $v$ separate times per problem. After generating $\mathbf{\psi}$ $v$ times, we select the median, conditioned on each gamma value, denoted $M(\psi_{\gamma_i v})$ among the $v$ samples as
\begin{align}
\psi_{\gamma_i} &= M(\psi_{\gamma_i v}),
\end{align}
where $\gamma_i$ corresponds to the $i^{th}$ ordered entry of the $\gamma$s sampled from the grid and therefore the new $i^{th}$ entry of $\mathbf{\psi}$. We note the algorithm does not always converge to the global minimum and at times provides unusable layouts. This limitation is also present in the standard FR algorithm and hence it is typically ran several times with the best plot being selected. This incidentally increases the value of our metric and biases the value upward. We select the median to be robust to these outliers. Furthermore, as a result of selecting the minimum value in our metric, we preference $\gamma$ values that provide more stable layouts. In this study, we select $v = 50$ to maintain a large enough sample size while balancing computational burden.
%Hence, choosing $\gamma$ that optimizes $\mathbf{\psi}$ will often correspond to the $\gamma$ at the beginning of the plateau in $\mathbf{m^S}$. However, we expect $\mathbf{m^S}$ to provide more consistent results in part due to the added variance in $\mathbf{\psi}$ and the redundant information located in $\mathbf{e^S}$ while noting the clarity explicitly modeling $\mathbf{e^S}$ brings. For the metric $\mathbf{m^S}$, we believe it is appropriate to select $\gamma$ = 1 when no plateau is present and the metric is monotone decreasing. We also recommend selecting $\gamma$ = 0 when the metric is constant in $\gamma$. Work still needs to be done on determining a plateau etc. 

\subsection{Simulation}

We consider different feature structures, network linkage probabilities, and simulations selecting $\gamma$ based on $\mathbf{\psi}$ and demonstrate the utility of our selection algorithm. We consider networks of sizes $N$ = 20, 50, 100, and 200. We control our linkage probabilities such that the average degree per node would be roughly 5, and investigate the distributions of the linkage probabilities such that the odds of connecting within group vs between groups are \{1:1, 1.5:1, 2:1, 2.5:1, 3:1, 3.5:1, 4:1, 4.5:1, 5:1\} among the participants. We integrate the odds of a connection being within group in categorical and continuous nodal covariates separately and only simulate the univariate case, although our results can be generalized. We do not account for missing network data here.

% Extended comment: should we separate this into *method* and *experiments demonstrating the method*? In the former, we could describe that we use logistic regression to estimate connection probabilities (and I'm shamelessly going to promote https://doi.org/10.1098/rsif.2020.0638 as an appropriate reference) and how we encode continuous and categorical features. In the latter, we can then apply the method to synthetic examples.

%\subsubsection{Categorical}
For categorical covariates, we generate these networks using a planted partition stochastic block model (SBM) with two and five groups. We fix the number of nodes and the number of groups. We then select a within group connection probability and a between group connection probability. Nodal covariates $X$ correspond to the group identification in the SBMs. We constrained probabilities of connections within groups and between groups by using a weighted average of potential links such that the average degree is maintained. We also selected the probabilities such that the within connection probability is up to 5x more likely than the between connection probability. 

%\subsubsection{Continuous}
For continuous covariates, we estimate the probability of connections between nodes $i$ and $j$, $B_{i j}$, as
\begin{equation}
    \hat{B}_{i j} = \mathrm{expit}\left(\beta_0 - log(\beta_1) \left\vert X_i - X_j\right\vert\right).
\end{equation}
Here, nodal covariates, $X$, are sampled from a uniform distribution $U(L, U)$. $\beta_1$ is the decrease in the odds of a connection between persons $i$ and $j$ associated with a one unit change between the distance of $X_i$ and $X_j$. $\beta_0$ is selected via optimization such that the average degree is maintained. $L$ and $U$ represent the lower and upper bound of the uniform distribution, respectively, fixed here at L = 0 and U = 1. We generate the network edges utilizing the corresponding probabilities. Note that as the difference in $X$ values increases, the likelihood of forming a connection decreases. So, in this case, nodes with similar $X$ values form connections with a higher probability. 

\subsection{Robustness to Missing Data}

The Procrustes transform is a mathematical technique used to align and compare two sets of points in a multidimensional space \citep{procrustes}. The transform works by finding an optimal transformation that minimizes the $L_2$ norm between the two sets of points. To perform the transform, the algorithm first centers the two sets of points by subtracting their respective means. This ensures that the transformed points are aligned around the origin of the coordinate system. Next, the algorithm scales the points to have unit variance, which normalizes the scale of the shapes. The transformed points are then aligned by finding a transformation that minimizes the sum of squared differences between corresponding points. The Procrustes aligns the points as closely as possible and allows researchers to compare shapes or data sets in a meaningful way by removing differences in position, scale, and orientation. We can then compare two network layouts using the Procrustes transform. The transform is needed because it helps with aligning graphs with similar structures, even if their node orderings differ. This is useful when comparing networks with similar topologies but different node identifiers. The nodal positions are considered identical if the loss function after utilizing the transform is 0 and completely different if the value is 1. To evaluate the robustness of our layout algorithm to edges being missing completely at random, we optimize the nodal positions when there is no missing edges as the reference layout and calculate the transform for the nodal positions at varying levels of missingness. Missing completely at random refers to a scenario in which the missingness of data is unrelated to both observed and unobserved variables, indicating that the missing data occurs randomly and independently of any factors or patterns within the dataset \citep{MCAR}. It is worth noting as the number of nodes converges to infinity the transform value converges to 1. Hence we only view a network with a 100 nodes. We compare our results to the standard FR algorithm.%For this section, we select for $\gamma$ and illustrate the general validity of our metric. We assume accurate nodal feature data. We also present many different literature methods to estimate $\mathbf{\hat{B}}$ however restrict our analysis to logistic regression and feature engineering. 
\section{Results}
\label{sec:results}

\subsection{Simulations}
\begin{figure}
\includegraphics[width = 0.8\textwidth]
        {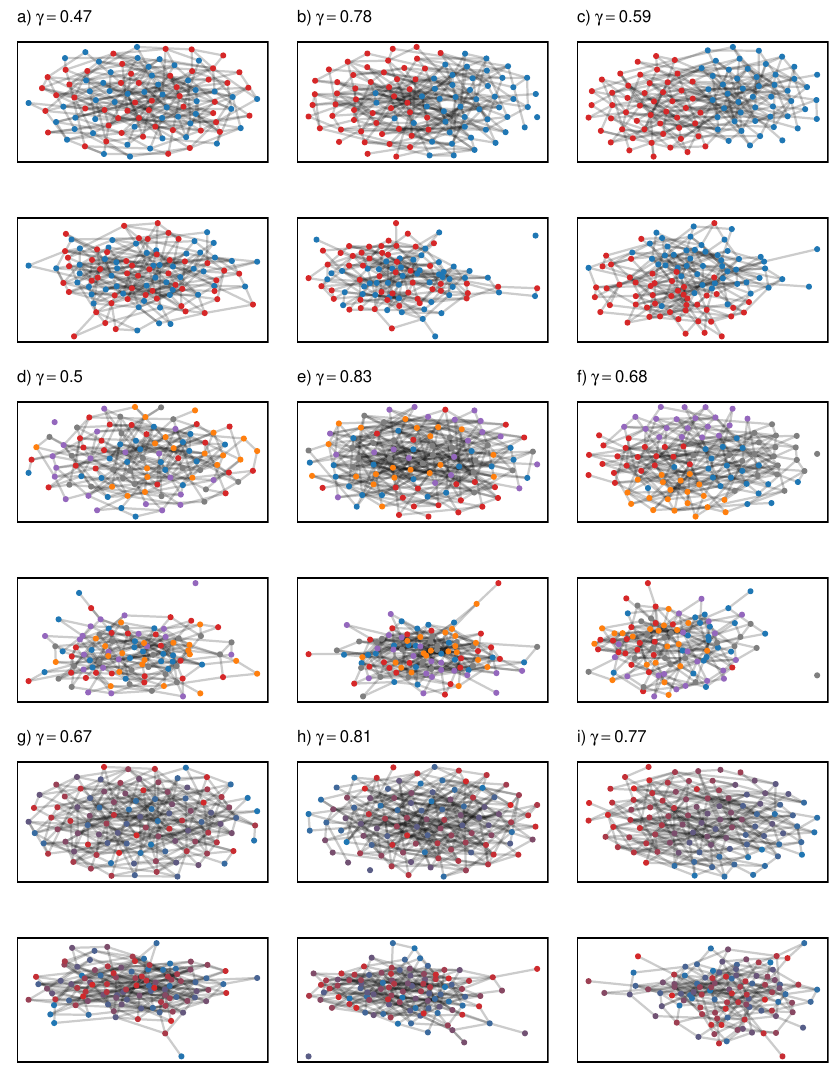}
\caption{Each two-row block corresponds to one of the three scenarios where nodal covariates have two categories (top), have five categories (middle), or are continuous (bottom). Each column represents odds: 1:1 (left), 1.5:1, 3.5:1 (right). For each scenario and odds combination, the visualization generated by our method is shown on top and the corresponding Fruchterman and Reingold (FR) visualization is shown on bottom. Each graph has 100 nodes, the  selected $\gamma$ value for our method is presented above each graph.}
\label{Selection_Plots}
\end{figure}

There are several ways to specify the dyadic model $M$ that predicts the linkage probability of a pair of nodes from their nodal covariates. For simplicity and interpretability, we restrict our analyses to logistic regression and assume accurate and fully observed nodal covariates in all analyses. We consider three scenarios: two scenarios involving categorical nodal covariates with two and five categories and one scenario involving continuous nodal covariates. In Figure \ref{Selection_Plots}, we visualize a 100-node network from each scenario using our method and the FR method, for odds that are 1:1, 1.5:1, and 3.5:1. The value of the tuning parameter $\gamma$ is shown in each panel and was selected using our selection algorithm described section \ref{subsec:tuning}.

In our algorithm, we combine information relating the nodal covariates with edge linkage probabilities and incorporate this information in our layout. When the odds are 1:1, nodal covariates carry no information about edge linkage, and there is no relevant information to incorporate in the model, we observe that there is no added clustering benefit to utilizing our algorithm. However, when the odds increase at a modest level to 1.5:1, we see that some scenarios benefit from smoothing while others do not. In particular, when there are two groups (categories), although the odds are modest, we see clear clustering on the basis of covariate values in panel b) of Figure \ref{Selection_Plots}. However, when there are five categories of nodal covariates, or the covariates are continuous as in panels e) and h), we only see low levels of clustering, although there still appears to be some benefit to our method. Finally, when we visualize higher odds at 3.5:1, there is a clear benefit from smoothing across all three scenarios. We also see that the two group case does not require as much smoothing as the other two cases which is reflected in its smaller $\gamma$ value. This is due to the signal strength of the odds. In general, higher signal strengths will require less smoothing, and lower nodal covariate complexity, more nodes, or high odds will increase said signal.

\begin{figure}
\includegraphics[height = 0.8\textwidth]
        {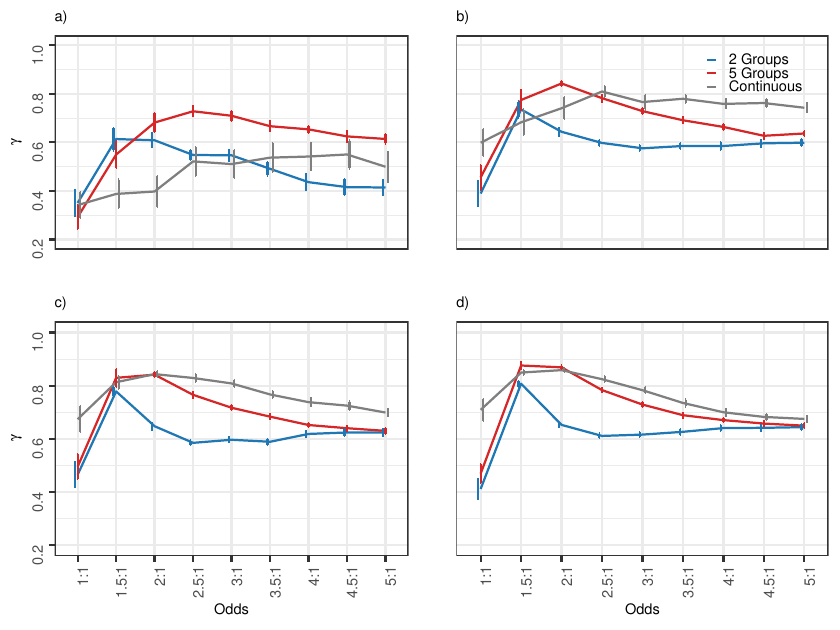}
\caption{Choosing an appropriate value for the smoothing parameter $\gamma$. We utilize our selection metric described in Section \ref{subsec:tuning} a total of 100 times per graph size, type of nodal covariate, and odds. We select $\gamma$ as the average among the 100 samples and present the 95\% confidence interval of the mean. We stratify our plots by graph size and data type. The number of nodes is a) 20, b) 50, c) 100, and d) 200 across the panels. The 5 Groups error bar is on the true odds while the other data types are offset.}
\label{Gamma_Trend}
\end{figure}

In Figure \ref{Gamma_Trend}, we see a trend such that as we increase the odds, the value of the selected smoothing parameter $\gamma$ increases to a point and then steadily decreases. Even odds contribute noise to the layout as the connection probabilities encourages our algorithm to randomly place the nodes, and our selection algorithm performs poorly in these scenarios. However, as we increase the odds, $m_{\gamma}$ begins to impact our selection and improves our results. We also see that the standard error for our selection decreases as the signal increases. 

\begin{figure}
\includegraphics[]
        {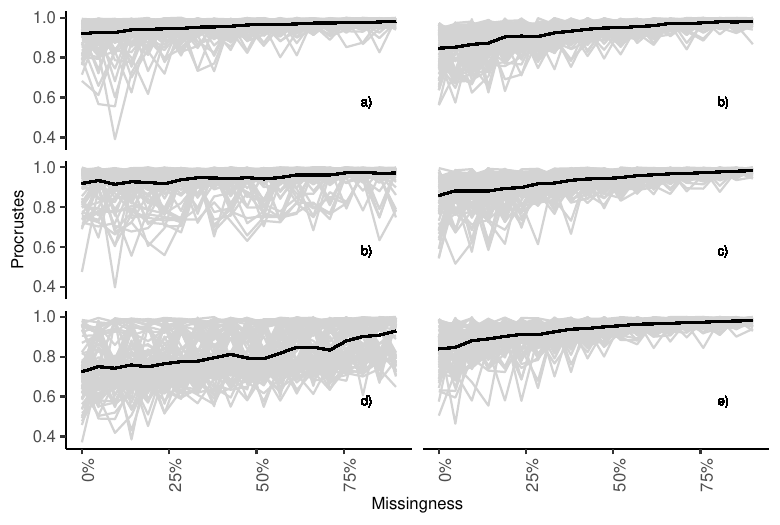}
\caption{Impact of missing edges on graph layout discrepancy. We plot the average Procrustes value between the nodal coordinates of a graph with a fixed percentage of missing edges and the nodal coordinates of a graph with no missing edges. The nodal coordinates are considered minimally different if the value of the Procrustes distance is 0 and maximally different if the value is 1. The grey lines represents the 100 realizations contributing  to the average (shown in black). Results are stratified by algorithm, our algorithm (left) and the Fruchterman-Reingold algorithm (right), and by odds: 1:1 (top), 1.5:1 (middle), and 3.5:1 (bottom). All plots represent a graph with a 100 nodes and one continuous nodal covariate sampled from the uniform distribution. The value of $\gamma$ is selected separately for each point in the graph. In general, our algorithm outperforms the FR algorithm when there are increased odds, whereas the outcome is reversed when the odds are even.}
\label{Procrustes}
\end{figure}

%For this section, we choose a graph and corresponding $\gamma$ to illustrate the general robustness to missing nodes and edges our layout provides compared to the standard FR algorithm. 
%\subsubsection{Results 2}
So far we have investigated graphs with no missing edges. If edges are missing, as is often the case, both the observed graph $G$ and the dyadic model $M$ are affected. Using a simulations, we study the impact of missing edges on the graph layout in three steps: First, we obtain a reference layout for the graph without missing edges. Second, we optimize graph layouts for varying levels of missing edges. Finally, we align the layouts for graphs with missing edges with the reference layout using the Procrustes transform and evaluate the Frobenius norm of residuals to quantify the departure from the reference layout, as shown in Figure \ref{Procrustes}. We see that the amount of information present in the covariates impacts the robustness of our algorithm. In panels a) and b), corresponding equal odds, the FR algorithm outperforms ours due to there being no relevant information in the covariates. As a consequence, attempting to smooth the layout ends up adding noise to the plot. However, in panels f) and g), corresponding to 3.5:1 odds, there is sufficient information in the covariates about edge linkage probabilities and our algorithm outperforms the FR.

\begin{figure}
\includegraphics[width=\textwidth]
        {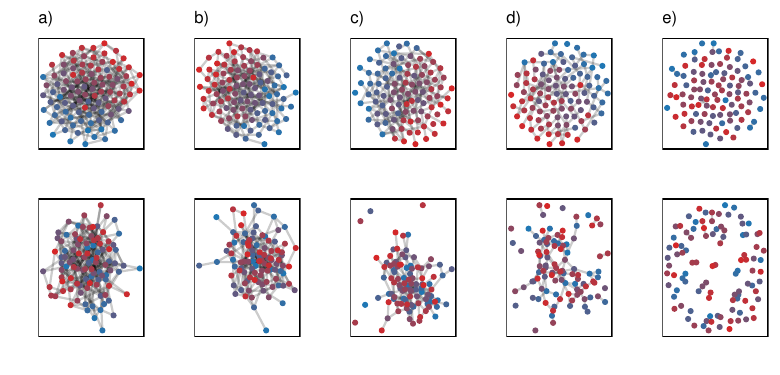}
\caption{Visualization of graphs with missing edges. We plot a graph with varying levels of missingness with our algorithm (top) and the FR algorithm (bottom). Proportion of missing edges ranges across the panels: a) 0\%, b) 22.5\%, c) 0.45\%, d) 67.5\%, and e) 90\%. All plots represent a graph with a 100 nodes and one continuous nodal covariate sampled from the uniform distribution with odds set at 3.5:1 and $\gamma$ is selected individually for each plot. In this high odds setting, the robustness of our algorithm appears to outperform the FR algorithm.}
\label{Missing_Vis}
\end{figure}

This phenomenon is more clearly demonstrated in Figure \ref{Missing_Vis}. Here we note the overall structure of our layout, when the odds is 3.5:1, is maintained when there is upward of 65\% edges missing completely at random in this example. 

\subsection{Data Application}
%Apply method to real world data
As an application of our network visualization method, we investigate friendship data from the Adolescent to Adult Health (Add Health) dataset on Community 4 \citep{moody}. The data are also available online \citep{tillahoffmann}. We construct two layouts: 1) incorporate all present covariate information (sex, race, grade, school) and 2) incorporate only the grade covariate. We refer to the former as the full model. In the Add Health study, every student received a paper-and-pencil questionnaire along with a roster containing the names of all students in the school. In cases where there were two schools in the community, students were provided with the roster of the corresponding ``sister'' school. The name generator solicited information on five male and five female friends separately. In our example, each grade is unique to a specific school. We placed an undirected edge between two students if either student named the other.

\begin{figure}
\includegraphics[width=\textwidth]
        {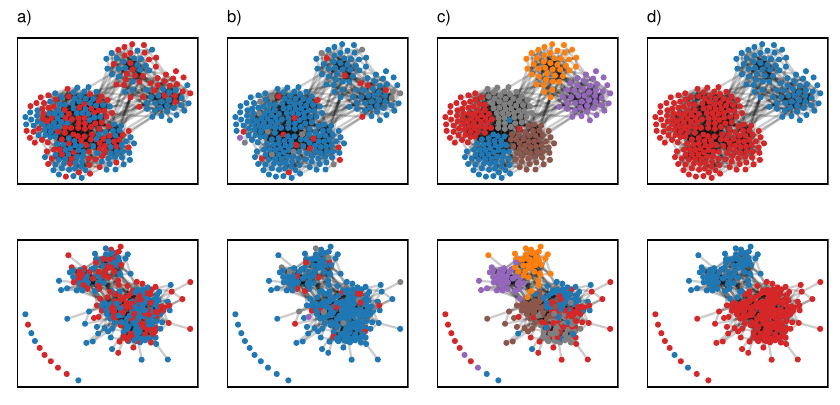}
\caption{Visualization of network data from the Add Health study using our method with the full linkage model (top row) and the FR algorithm (bottom row). All pairwise linkage probabilities were estimated using logistic regression, and all covariates are 1 if two nodes share the category and 0 if not. We considered four covariates: a) $\tilde{sex}$, b) $\tilde{race}$, c) $\tilde{grade}$, and d) $\tilde{school}$. The coefficient estimates  and standard errors are as follows: $\tilde{sex}$ 0.14 (0.062), $\tilde{race}$ 0.11 (0.079), $\tilde{grade}$ 1.99 (0.070), and $\tilde{school}$ 1.92 (0.163). We chose $\gamma$ = 0.672 using the procedure described in the paper. We colored the nodes in each panel by the values of the corresponding categorical covariate.}
\label{Comm_Vis}
\end{figure}

We estimated pairwise linkage probabilities in both models using logistic regression which included an intercept term. The final estimated full model was the following:

\begin{align}
    \log \left(\frac{P(A_{ij} = 1)}{1 - P(A_{ij} = 1)} \right) &= -6.15 + 0.14\times\tilde{sex}_{ij}  + 0.11\times\tilde{race}_{ij} + 
    1.99\times\tilde{grade}_{ij} +
    1.92\times\tilde{school}_{ij},
\end{align}

where $\tilde{x}_{ij}$ = 1 if nodes $i$ and $j$ have the same value of $x$ and 0 otherwise. In Figure, \ref{Comm_Vis} we see that our algorithm has similarities to FR. We see these similarities due to the observed covariates capturing the majority of the variance in edge formation and the presence of high odds exceeding 5:1 in these data when two students are only at the same school together compared to having nothing in common. We note both school and grade level largely impact edge formation, and in our layout the schools are placed at a greater distance apart than in FR. We also see singleton nodes that are placed mostly with their respective groups while the FR places them on the outskirts of the panel. Finally, it appears clear that our algorithm segregates the different groups more clearly.

The final estimated grade-only model was the following: 
\begin{align}
    log \left(\frac{P(A_{ij} = 1)}{1 - P(A_{ij} = 1)} \right) &= -5.99 + 
    2.91\times\tilde{grade}_{ij}. 
\end{align}

\begin{figure}

\includegraphics[]{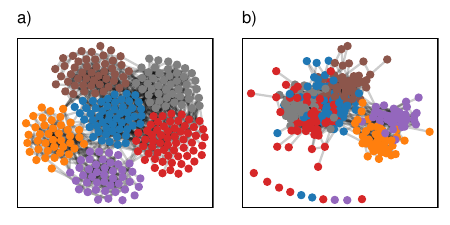}
\caption{Visualization of network data from the Add Health study using (a) our method with the grade-only linkage model and (b) the FR algorithm. The coefficient estimate and standard error is as follows: $\tilde{grade}$ 2.91 (0.156). Utilizing our selection procedure, we choose $\gamma$ = 0.589. We colored the notes in each panel by grade.}
\label{Comm_Vis_grade}
\end{figure}

When only considering the grade of the children in Figure \ref{Comm_Vis_grade}, we still observe the purple and orange grades separated from the other grades due to them uniquely corresponding to the second school. However, these two group are no longer distantly separated from the other grades. We limited our univariate analysis to only the grade due to its strong correlation with school and the low effect size among sex, and race, but note any of the covariates could be used to draw a unique layout.

\section{Conclusion}
\label{sec:conclusion}

We envision that our method is especially powerful in a situation where three conditions are met: we have reliable nodal covariates, we have unreliable accounts of network structure, and we expect nodal covariates to be highly predictive of node linkage probability. In this study, we see that when there is relevant information on edge connections contained inside nodal covariates, there is an added visual benefit to placing a smoothing component within the standard FR algorithm that incorporates the relevant information in the layout. Furthermore, including this information in the layout provides robustness to edges missing completely at random when all nodal information is accurate. Also, we recommend methodology to carefully select the tuning parameter for smoothing that provides optional aesthetic features such as shorter edges lengths. Finally, we note our algorithm can smooth a layout relative to any combination of covariates. We acknowledge our algorithm contributes noise to FR when there is not relevant information on edge connections within the covariate(s), and recommend the standard FR in these scenarios. 

\section{Reproducibility statement}
The code for reproducing this paper is available at \url{https://github.com/onnela-lab/covariate-smoothed-layout}.

\section{Funding details}
The authors were supported by NIH/NIAID grant R01AI138901.

\section{Disclosure statement}
The authors have no conflict of interest.

% reproducibility statement is mandatory for all papers
%\begin{reproduce}
%The code for reproducing this paper is available at \%https://github.com/onnela-lab/covariate-%smoothed-layout
%\end{reproduce}

\bibliographystyle{plainnat}
\bibliography{reference}

\end{document}